\documentclass[preprint,aps,draft]{revtex4}
\usepackage{epsf}
\begin{document}
\title{Propagation gap for shear waves in binary liquids: Analytical and simulation study
}

\author{Taras Bryk$^{1,2}$, Maria Kopcha$^{1}$}

\affiliation{ $^1$ Institute for Condensed Matter Physics of NAS 
 of Ukraine,\\UA-79011 Lviv, Ukraine}
\affiliation{$^2$Institute of Applied Mathematics and Fundamental
Sciences,\\Lviv National Polytechnic University, UA-79013 Lviv,
Ukraine} 

\date{\today}
\begin{abstract}
Transverse collective excitations in a binary liquid mixture are studied for different mass ratio 
of components $R$ at fixed numerical density. Increasing the mass ratio results in a larger gap 
width for shear waves. A four-variable dynamic model of transverse dynamics in binary liquids is 
solved analytically with an account of cross-correlations 
between total mass and mass-concentration current fluctuations. An equation for 
the propagation gap of shear waves for binary liquids is reported and analyzed.
\end{abstract}

\maketitle

\section{Introduction}
Transverse dynamics in liquids is very simple on macroscopic scales, where the only shear 
relaxation perfectly describes the long-time behavior of wave-number-dependent transverse 
mass-current 
autocorrelations for sufficiently small wave numbers $k$\cite{Boo,Han}:
\begin{equation}\label{hydT}
F^T_{JJ}(k,t)\stackrel{k\to 0}{=}mk_BTe^{-D_vk^2t}
\end{equation}
with $D_v=\eta/\rho$ being kinematic viscosity, $\eta$ - shear viscosity, and $\rho$ - mass density. 
This fact is a 
consequence of existing the only transverse conserved quantity - transverse component 
of total momentum - that excludes any transverse wave-like motion on macroscopic scale in liquid, 
i.e. explains the absence of transverse sound in liquid systems. From (\ref{hydT}) it follows, that
 the hydrodynamic long-time behavior of shear relaxation for any liquid 
(simple or complex) will be qualitatively similar. Outside the hydrodynamic regime, with increasing
wave numbers, the overdamped propagating transverse excitations called shear waves can emerge in 
the liquid. Shear
waves can propagate only on the nanoscales because of the strong damping and this fact makes them 
absolutely different from the transverse sound in crystalline and glassy solids. Inability of 
the shear waves to exist on macroscopic distances is reflected in their dispersion law $\omega(k)$,
which was obtained in the end of 1960-s\cite{Chu69,Boo}, and makes evidence of the propagation 
gap for shear waves at small wave numbers. Later, intensive studies by molecular dynamics (MD) 
simulations 
theory of non-hydrodynamic shear waves were focused on damping effects and different aspects 
of transverse dynamics of liquids\cite{Ald84,Mac84,Bry00,Bry11}. The developed theory 
\cite{Mac84,Bry00,Bry11} explicitly connected the emergence of shear waves and their damping 
with non-hydrodynamic shear-stress relaxation. Surprisingly, in 2017 Trachenko {\it et al} 
\cite{Yan17} obtained from a macroscopic continuum equation
the same expression for the propagation gap of shear waves, and claimed 
that ill-defined "Frenkel jumps" and their characteristic {\it single-particle} 
"Frenkel time" are responsible for the propagation gap, that was immediately 
criticized in\cite{Bry18}. 

For many-component liquids the list of non-hydrodynamic collective modes is much larger, because 
typical high-frequency overdamped excitations for binary and many-component liquids are optic-like 
collective modes \cite{Bry99,Bry00,Bry02,Bry11}. 
Generalized hydrodynamics for the case of transverse dynamics
in binary liquids was developed in \cite{Bry00b}, and in \cite{Bry00} 
were derived expressions for the dispersion and damping of transverse optic-like modes in the 
long-wavelength limit within the formalism of generalized collective modes (GCM) \cite{Mry95}. 
It was revealed that the mutual diffusivity and tendency to demixing (large 
values of the concentration-concentration structure factor $S_{xx}(k\to 0)$) define the damping and
can cause a similar propagation gap in $k$-space in case of larger value than the critical damping 
for optic modes. And it became quite clear why for ionic liquids the optic modes are well defined - 
because of $S_{xx}\stackrel{k\to 0}{\to}0$ due to electroneutrality condition practically vanishing
damping of optic modes in the long-wavelength limit was observed for molten salts\cite{Bry04}. The dispersion 
curves of longitudinal and transverse excitations for binary liquids were intensively studied 
within the GCM methodology\cite{Bry00,Bry02,Bry04,Mry09}, however, the issue of the width of propagating gap 
for shear waves in binary and many-component liquids was not studied analytically. There exists just the 
standard propagation gap width \cite{Mac84,Bry00}
\begin{equation}\label{k-gap}
k_s=\frac{\sqrt{\rho G_{\infty}}}{2\eta}~,
\end{equation}
where $G_{\infty}$ is the macroscopic high-frequency shear modulus. However, the shear waves as non-hydrodynamic 
excitations couple with optic modes in many-component liquids and this coupling should affect the location of 
$k_s$. In order to obtain the analytical expression for the gap width in this
case one has to solve analytically in the long-wavelength limit a four-variable dynamic model, 
which incorporates two two-variable dynamic models solved in \cite{Bry00} separately. 
The numerical GCM approach for the gap width for shear waves in binary liquids points out 
an increase of the gap width with the mass component ratio, while an increase of concentration of 
heavier particles shows the opposite tendency\cite{Mry09}. Therefore it will be of great interest 
to obtain analytical expressions to explain the simulations and GCM numerical studies.

The remaining part of the paper is organized as follows. In the next Section we 
report our MD simulations of four binary liquids with different mass ratio, which
however have identical static structure, and numerical GCM results for dispersions 
oftwo branches of transverse collective excitations.
In Section III we formulate the 
four-variable dynamic model for transverse dynamics of a binary liquid, generate a generalized 
hydrodynamic matrix for it and obtain four dynamic eigenmodes in the long-wavelength limit as
well as an expression for the gap width $k_s$ for shear waves.
The last Section contains conclusion of this study.

\section{Computer simulations of transverse dynamics in a Kob-Andersen liquid with different 
mass ratio}

The idea of this study is to rationalize an effect of the mass ratio increase on 
transverse dynamics in binary liquids, and on the propagation gap for shear waves in 
particular. For longitudinal dynamics in binary liquids such an effect was studied 
in \cite{Bry05}, while for the case of transverse dynamics Mryglod {\it et al} \cite{Mry09} 
reported observation of the increasing gap width for shear waves vs. mass ratio, although
no theory explaining that finding was developed. Here we will study a binary liquids with 
identical static structure but different mass ratio of components $R=m_B/m_A$, will obtain the 
propagation gap dependence on $R$ and develop a theory able to explain the observed effect.

\begin{figure}
\epsfxsize=.60\textwidth {\epsffile{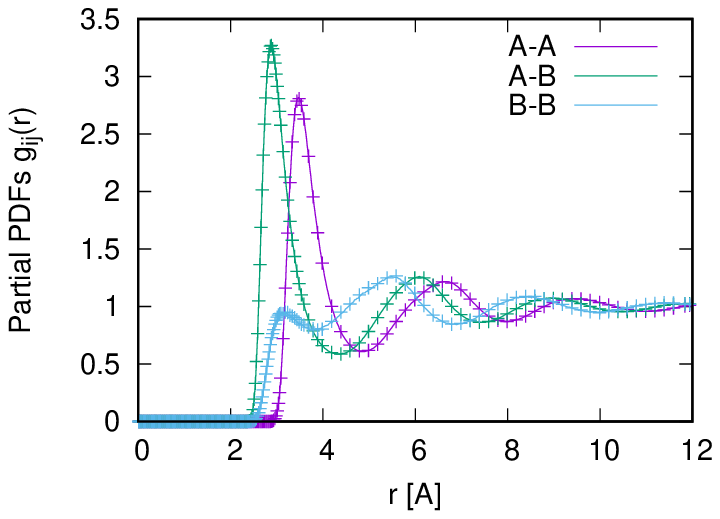}}
\caption{Identical static structure from pair distribution functions 
for Kob-Andersen liquid mixture at density $n^*=1.20$
and temperature $T^*=2.0$ with different mass ratio: R$=1$ (lines) and R$=20$ (plus symbols).
}
\label{Fig_pdf}
\end{figure}

First we simulated the standard Kob-Andersen 80-20 binary liquid with mass ratio 
$R=1$ \cite{Kob95,Ped18} at temperature $T^*=2.0$ and density $n^*=1.2$. The standard set 
of parameters for Lennard-Jones potential of the KA modes was used (in relative units to A-A 
interaction):
$\varepsilon_{AA}^*=1.0$, 
$\varepsilon_{AB}^*=1.5$, 
$\varepsilon_{BB}^*=0.5$,
$\sigma_{AA}^*=1.0$, $\sigma_{AB}^*=0.8$, $\sigma_{BB}^*=0.88$. All the output quantities are 
in dimensions of SI units. For simulations we have chosen $\varepsilon_{AA}=119.8~K$, 
$\sigma_{AA}=3.405$~\AA and ${\bar m}=7.960\cdot 10^{-26} kg$. We have fixed the total mass density 
of the system and generated starting configurations with additional 3 mass ratio 
 $R=2$, $R=8$, $R=20$, then set the total momentum of each starting configuration to zero 
and equilibrated each system in microcanonical ensemble till the average temperature was fluctuating
around $T^*$. Note, that prepared in this way the 4 simulated system have identical static 
structure but different dynamics. One can see in Fig.\ref{Fig_pdf} that there is no difference
between three partial pair distribution functions shown for the case of mass ratio $R=1$ and 
$R=20$ (for others, $R=2$ and $R=8$, this is the same).

For each system with corresponding mass ratio we performed production runs of 600 000 time steps
and saved time evolution of the following dynamic variables:
spatial Fourier-components of partial transverse mass-currents
\begin{equation}\label{Ji}
{\bf J}^{T}_{\alpha}(k,t)=\frac{1}{\sqrt{N}}
\sum_{j=1}^{N_{\alpha}}m_{\alpha}{\bf v}^{T}_{j,\alpha}(t)
 e^{-i{\bf kr}_{j,\alpha}(t)}~,\alpha=A,B,
\end{equation}
where ${\bf v}^{T}_{j,\alpha}(t)=[{\bf k}\times {\bf v}_{j,\alpha}]/k$ 
is the transverse projection (transverse component) of 
velocity ${\bf v}_{j,{\alpha}}$ of $j$-th particle of kind $\alpha$,
and their first time derivatives
\begin{equation}\label{dJLk}
\frac{d {\bf J}^T_{\alpha}(k,t)}{dt}\equiv {\dot {\bf J}}^T_{\alpha}(k,t)
=\frac{1}{\sqrt{N}}\sum_{j=1}^{N_{\alpha}}
[{\bf F}^T_{j,\alpha}(t)-im_{\alpha}({\bf kv}_{j,\alpha})
{\bf v}^T_{j,{\alpha}}(t)]e^{-i{\bf kr}_{j,\alpha}(t)}~,
\end{equation}
where ${\bf F}^T_{j,\alpha}(t)$ is the transverse component of force 
acting on the $j$-th particle of kind $\alpha$. The four partial dynamic 
variables form a basis set 
\begin{equation}\label{Aij}
A^{(4Tp)}(k,t)=\{{\bf J}^T_{A}(k,t), {\bf J}^T_{B}(k,t), 
{\dot {\bf J}}^T_{A}(k,t), {\dot {\bf J}}^T_{B}(k,t)\}~,
\end{equation}
which is used for construction of the $4\times 4$ generalized hydrodynamic 
matrix for transverse dynamics of a binary liquid \cite{Bry00b}. All the 
matrix elements (static averages or correlation times) were
calculated directly from MD data. All the $k$-dependent quantities
were averaged over all possible directions of the ${\bf k}$ vectors with
the same absolute value. 

\begin{figure}
\epsfxsize=.60\textwidth {\epsffile{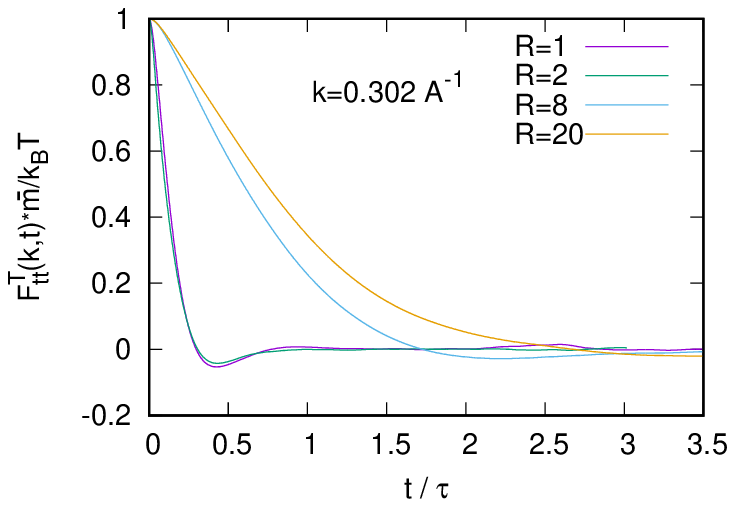}}
\caption{Transverse total current autocorrelation function at $k=0.302$\AA$^{-1}$
for Kob-Andersen liquid mixture with different mass ratio R$=m_B/m_A$ at $T^*=2.0$.
The timescale $\tau$ is 3.97299 ps.
}
\label{Fig_Ftt}
\end{figure}

The transverse total current autocorrelation functions $F^T_{J_tJ_t}(k,t)$ for 
binary liquids at fixed density and temperature but with different mass ratio have 
different relaxation, as it is shown in Fig.\ref{Fig_Ftt} at $k=0.302$\AA$^{-1}$.
For $R=1$ and $R=2$ the relaxation shape of $F^T_{J_tJ_t}(k,t)$ is typical like 
in the $k$-region with propagating shear waves, while for mass ratio $R=8$ and 
$R=20$ the shape is close to the hydrodynamic one (\ref{hydT}).

\begin{figure}
\epsfxsize=.45\textwidth {\epsffile{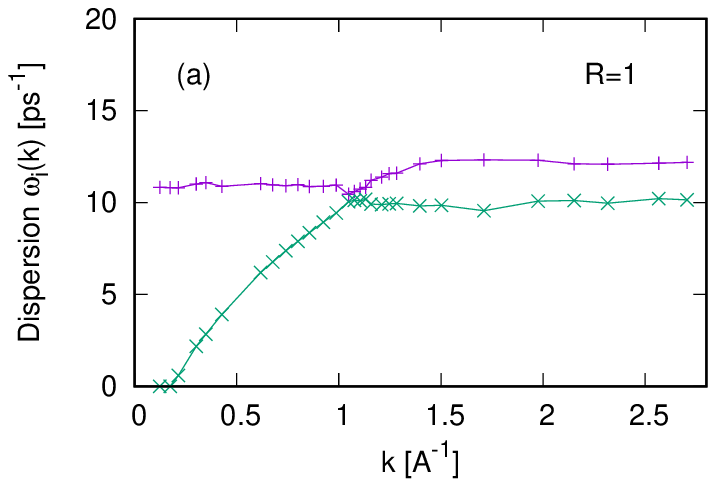}}
\epsfxsize=.45\textwidth {\epsffile{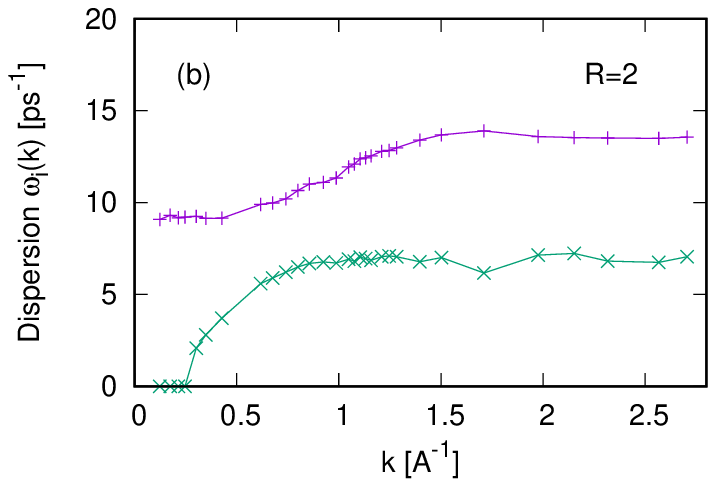}}

\epsfxsize=.45\textwidth {\epsffile{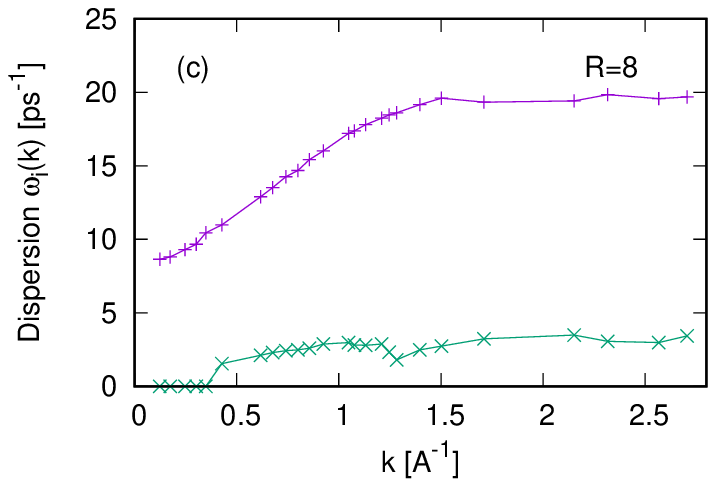}}
\epsfxsize=.45\textwidth {\epsffile{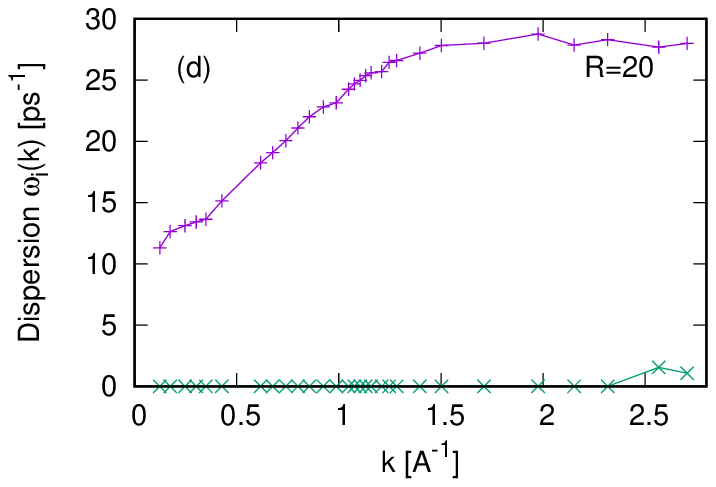}}
\caption{Dispersion of transverse optic and shear wave eigenmodes, obtained by GCM method,
 for Kob-Andersen liquid mixture with different mass ratio R at $T^*=2.0$.
}
\label{Fig_wk}
\end{figure}

This finding stimulated us to calculate the dispersion curves of GCM propagating eigenmodes
for binary liquids at fixed density but  with different mass ratio R
using the $4\times 4$ generalized hydrodynamic matrices generated on the 4-variable set
(\ref{Aij}). In Fig.\ref{Fig_wk} we show the dispersion of two branches of transverse propagating 
modes: low-frequency shear waves and high-frequency optic-like modes (for which the nearest 
neighbors move in opposite directions in transverse planes with respect to the propagation 
vector ${\bf k}$). The most striking feature in Fig.\ref{Fig_wk} is the strong effect of the 
mass ratio on the low-frequency branch of shear waves: the width of their propagation gap 
increases with $R$ and for large mass ratio of components one can expect the low-frequency
branch of shear waves to disappear totally from the frequency spectrum.

\begin{figure}
\epsfxsize=.60\textwidth {\epsffile{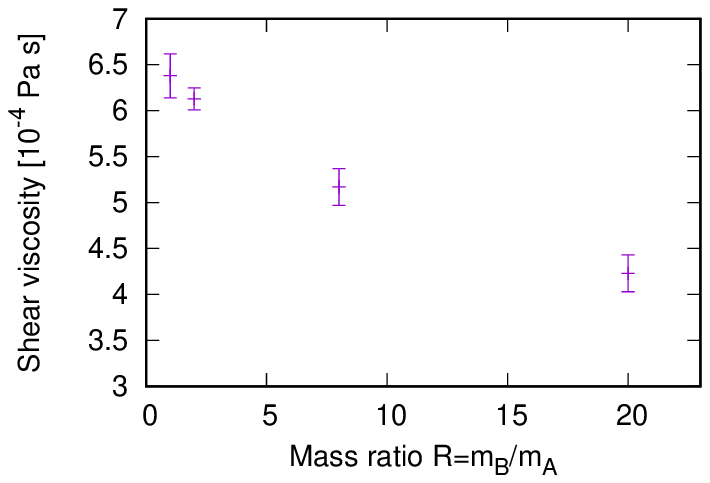}}
\caption{Shear viscosity as a function of mass ratio R, obtained via Kubo intergrals of
shear-stress autocorrelation functios for Kob-Andersen liquid mixture at $n^*=1.2$ and 
$T^*=2.0$.
}
\label{Fig_eta}
\end{figure}

In order to check how these findings agree with the expression for the gap width for 
shear waves (\ref{k-gap}) we calculated the shear-stress autocorrelation functions 
and estimated shear viscosity for the 4 simulated system with fixed density but different 
mass ratio. As expected, the identical static structure resulted in identical (within 
error bars) values of the high-frequency shear modulus $G_{\infty}=3.36\pm 0.02~GPa$. 
The shear viscosity, though, was obtained from the standard Kubo integration of the   
shear-stress autocorrelation functions and showed a small decrease with increasing 
mass ratio $R$ (Fig.\ref{Fig_eta}). In \cite{Bry05} longitudinal collective dynamics 
was studied for equimolar binary liquids with fixed mass density and different mass ratio,
and it was observed the increase of mutual diffusivity vs mass ratio $R$. Hence, it is 
consistent with the actual observation that the shear viscosity was decreasing with
$R$ as it is shown in Fig.\ref{Fig_eta}.

\begin{figure}
\epsfxsize=.60\textwidth {\epsffile{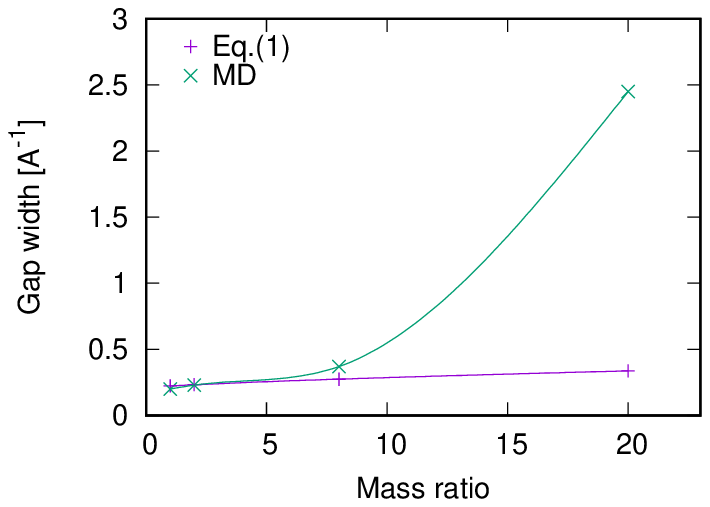}}
\caption{Gap width for shear waves vs. mass ratio R for a binary Kob-Andersen 
liquid with density $n^*=1.2$ at $T^*=2.0$ as obtained from Eq.1 
(plus symbols) and observed from MD simulations (cross symbols).
}
\label{Fig_gap}
\end{figure}

Now, having the values of $\eta$ and $G_{\infty}$ we can estimate how the width of
propagation gap for shear waves changes according to Eq. (\ref{k-gap}) and from 
MD simulations. In Fig.\ref{Fig_gap} we show the both dependences,and it is seen that 
the expression for gap width is not working for binary liquids, especially with mass
ratio larger than 2. Hence, there is a need to develop a theory for shear waves 
in binary liquids accounting for the coupling between two high- and low-frequency 
branches of collective 
excitations (i.e. cross-correlations between total and mass-concentration fluctuations).

\begin{figure}
\epsfxsize=.60\textwidth {\epsffile{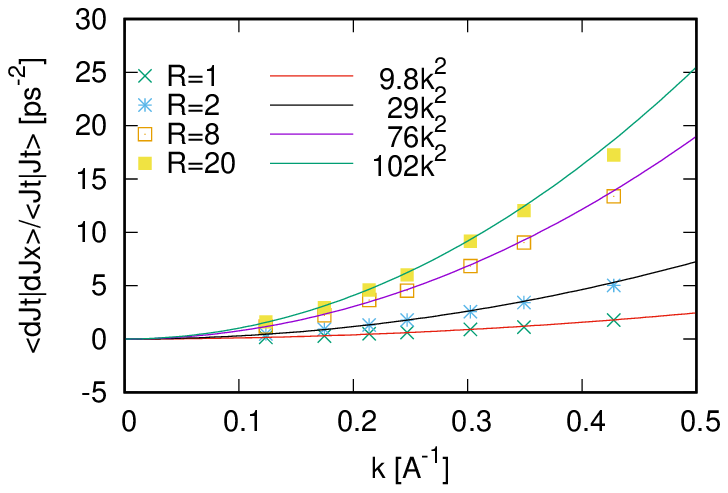}}
\caption{Cross-correlations $\langle{\dot J}^T_t(-k){\dot J}^T_x(k)\rangle/
\langle J^T_t(-k) J^T_t(k)\rangle$ as a function of wave number at different
mass ratio R. Symbols - MD values, fitting solid lines - evidence of the 
$k^2$-dependence in the long-wavelength region with increasing coupling 
constant $a$ vs R.
}
\label{Fig_djdj}
\end{figure}

The static cross-correlation between the transverse total and mass-concentration currents
is zero as it was shown in \cite{Bry00}. However, the cross-correlation exists between 
their first time derivatives, and we can directly calculate them from MD simulations.
In Fig.\ref{Fig_djdj} we show the evidence that the directly estimated from MD
cross-correlations 
$\langle{\dot J}^T_t(-k){\dot J}^T_x(k)\rangle$ are proportional in the long-wavelength 
limit to $k^2$, and increase with increasing mass ratio $R$. Having this information we 
will try to solve analytically a  4-variable dynamic model for description of the 
transverse dynamics in binary liquids in order to derive the equation for the propagation 
gap $k_s$ of shear waves.

\section{Four-variable theory of generalized collective modes for transverse dynamics}

We start from the choice of a set of dynamic variables for defining the generalized hydrodynamic 
matrix ${\bf T}(k)$ for analytical study of transverse dynamics of binary liquids. The eigenmodes of ${\bf T}(k)$ 
allow to represent the time correlation functions of interest via a weighted sum of mode 
contributions in a wide range of wave numbers $k$\cite{Mry95}:
\begin{equation}\label{Fkt}
F^T_{JJ}(k,t)=\sum_{i=1}^{N_v}G_{JJ}^i(k)e^{-z_i(k)t}~,
\end{equation}
where $z_i(k)$ are the (complex in general case) eigenvalues of the $N_v\times N_v$ matrix 
${\bf T}(k)$ for the transverse dynamics, and the weight coefficients (complex in general case)
$G_{JJ}^i(k)$ are estimated via corresponding eigenvectors \cite{Mry95}.
The simplest choice of dynamic variables for transverse dynamics would be to 
start from partial
currents of A and B components (\ref{Ji}), as it is quite simply to sample them in MD simulations
and use in numerical GCM study. Note that the 
GCM eigenvalues will be identical for matrices ${\bf T}(k)$ generated either 
on a set of partial dynamic 
variables or on a set of any their linear combination. For analytical study it is better to start 
from hydrodynamic variables which correspond to fluctuations of the conserved quantities. Therefore
the hydrodynamic variable of transverse component of total mass current 
\begin{equation}\label{Jt}
{\bf J}^{T}_{t}(k,t)=
 \sum_{\alpha}{\bf J}^{T}_{\alpha}(k,t)~,\alpha=A,B,
\end{equation}
is simply the sum of partial currents $J^{T}_{\alpha}(k,t)$, and one can form from the 
partial currents another dynamic variable of mass-concentration current
\begin{equation}\label{Jx}
{\bf J}^{T}_{x}(k,t)=x_B{\bf J}^{T}_{A}(k,t)-x_A{\bf J}^{T}_{B}(k,t)~,
\end{equation}
which is orthogonal to the hydrodynamic variable, i.e.
$$\langle {\bf J}_x^T(-k){\bf J}_t^T(k)\rangle\equiv 0~. $$
In (\ref{Jx}) $x_A=m_AN_A/(m_AN_A+m_BN_B)$ and $x_B=1-x_A$ are mass-concentrations.
Having two orthogonal dynamic variables (\ref{Jt}) and (\ref{Jx}) we can extend the 
inital set of these two ones by other orthogonal dynamic variables, constructed as it is standard
in the GCM approach via first time derivatives of (\ref{Jt}) and (\ref{Jx}). Hence, we 
have obtained a set of four variables 
\begin{equation}\label{basis4}
A^{(4T)}(k,t)=\{{\bf J}^T_{t}(k,t), {\bf J}^T_{x}(k,t), 
{\dot {\bf J}}^T_{t}(k,t), {\dot {\bf J}}^T_{x}(k,t)\}~,
\end{equation}
which will be used for generating the generalized hydrodynamic matrix for the case of 
transverse dynamics in binary liquids. Note, that in\cite{Bry00} the analytical solutions 
were obtained for two separated (uncoupled) 2-variable dynamic models:
for description of dynamics of total (t) mass-current fluctuations
\begin{equation}\label{basis2t}
A^{(2Tt)}(k,t)=\{{\bf J}^T_{t}(k,t), {\dot {\bf J}}^T_{t}(k,t)\}
\end{equation}
and for mass-concentration (x) current fluctuations
\begin{equation}\label{basis2x}
A^{(2Tx)}(k,t)=\{{\bf J}^T_{x}(k,t), {\dot {\bf J}}^T_{x}(k,t)\}~.
\end{equation}
The two-variable dynamic model (\ref{basis2t}) is equivalent to description of the 
transverse dynamics of one-component liquid and results in the long-wavelength limit 
in two relaxing (purely real eigenvalues) modes: the standard hydrodynamic mode (\ref{hydT}) 
\begin{equation}\label{z1}
z_1(k)\equiv Re\{z_1(k)\}=D_vk^2
\end{equation}
and a non-hydrodynamic shear-stress relaxation mode \cite{Bry11} with the purely real eigenvalue
\begin{equation}\label{z2}
z_2(k)\equiv Re\{z_2(k)\}=d_0-D_vk^2
\end{equation}
with 
\begin{equation}\label{d0}
d_0=\frac{G_{\infty}}{\eta}=\tau_M^{-1}
\end{equation}
where $\tau_M$ is the standard Maxwell relaxation time. The dispersion of shear waves \cite{Chu69}
which emerge outside the hydrodynamic regime of pure liquids is as follows:
\begin{equation}\label{z_shear}
z^{\pm}_{sw}(k)=\sigma_{sw}(k)\pm\sqrt{\langle \omega^2_{tt}\rangle(k)-\sigma^2_{sw}(k)},
\end{equation}
with the normalized second frequency momentum of total transverse mass-current spectral function 
$\langle \omega^2_{tt}\rangle(k)=\frac{k^2G_{\infty}(k)}{\rho}$ and damping $\sigma_{sw}(k)$, 
where 
$$c_T^2=\frac{G_{\infty}}{\rho}$$ is the square of propagation speed for macroscopic "bare" 
(non-damped) transverse excitations. Note, that $c_T$ depends on the total mass density and 
high-frequency shear modulus, and does not depend explicitly on the ratio of mass components
in the mixture.

Let us apply the set of four dynamic variables (\ref{basis4}) to construction of the 
$4\times 4$ generalized hydrodynamic matrix ${\bf T}^{(4T)}(k)$ following the methodology 
suggested in \cite{Mry95}. We result in 
\begin{equation}
\begin{array}{l}
        {\bf T}^{(4T)}
        (k)= \left(
\begin{array}{cccc}
  0    &     0    & -1                   &    0                 \\
  0    &     0    &  0                   &   -1                 \\
  c_T^2k^2    &    ak^2    & \frac{c^2_T}{D_v}   & \frac{ak^2 D_{12}S_{xx}}{x_1x_2k_BT}  \\
  x_1x_2ak^2    &   \langle\omega_{xx}^2\rangle &\frac{x_1x_2a}{D_v}&  
             \frac{\langle\omega_{xx}^2\rangle D_{12}S_{xx}}{x_1x_2k_BT}      \\
\end{array}
\right)\ , \label{Thyd}
\end{array}
\end{equation}
where the cross-correlations between the first time derivatives of total and 
mass-concentration currents was represented as
$\langle\dot{J}_x^T(-k)\dot{J}_t^T(k)\rangle=x_1x_2\bar{m}k_BTak^2$
with a coupling constant $a$.
Note, that when $a=0$ our $4\times 4$ generalized hydrodynamic matrix decomposes into two 
$2\times 2$ independent blocks representing separately dynamics of total transverse mass-current
\begin{equation}
\begin{array}{l}
        {\bf T}^{(2Tt)}
        (k)= \left(
\begin{array}{cc}
  0    &    -1       \\
  c_T^2k^2    & \frac{c^2_T}{D_v}  \\
\end{array}
\right)\ , \label{Thyd2}
\end{array}
\end{equation}
and mass-concentration transverse current 
\begin{equation}
\begin{array}{l}
        {\bf T}^{(2Tx)}
        (k)= \left(
\begin{array}{cc}
  0    &    -1       \\
  \langle\omega_{xx}^2\rangle &  
             \frac{\langle\omega_{xx}^2\rangle D_{12}S_{xx}}{x_1x_2k_BT}      \\
\end{array}
\right)\ , \label{T2x}
\end{array}
\end{equation}
within a precision of the second frequency moment of corresponding transverse current 
spectral function. Analytical solutions for eigenmodes of matrices (\ref{Thyd2}) and (\ref{T2x})
were reported in \cite{Bry00}. $\langle\omega_{xx}^2\rangle$ is the second frequency moment of 
mass-concentration transverse current spectral function.

For the $(4\times 4)$ eigenvalue problem with the generalized hydrodynamic matrix (\ref{Thyd})
we are trying to obtain the long-wavelength eigenvalues in a form of expansion in $k$ restricting 
the highest order of $O(k^2)$. Hence we obtained two purely real eigenvalues
\begin{equation}\label{z12rel}
\begin{array}{l}
z^{(4T)}_1(k)\equiv Re{z^{(4T)}_1(k)}=D_vk^2\\
z^{(4T)}_2(k)\equiv Re{z^{(4T)}_2(k)}=d_0-D_vk^2-\Delta a^2 k^2~.
\end{array}
\end{equation}
It is important to note, that the hydrodynamic relaxation mode $z^{(4T)}_1(k)$ is not affected 
by the non-hydrodynamic processes as it should be. The other two eigenvalues correspond to
 a pair of propagating eigenmodes $z^{(4T)}_{3,4}(k)\equiv z_{opt}^{\pm}(k)$, where
\begin{equation}\label{zopt}
\begin{array}{l}
z_{opt}^{\pm}(k)=\sigma_{opt}(k)\pm i\omega_{opt}(k)\\
\sigma_{opt}(k)=\sigma_0+\frac{\Delta}{2}a^2 k^2\\
\omega_{opt}(k)=\sqrt{\langle\omega_{xx}^2\rangle_{k}-\sigma_{opt}^2(k)}~,
\end{array}
\end{equation}
which are the optic-like modes. In (\ref{z12rel}) and (\ref{zopt}) we made a
 shortcut $\Delta$ for 
\begin{equation}\label{Delta}
\Delta=\frac{x_1x_2}{D_v}\frac{1-\frac{2\sigma_0d_0}{\langle\omega_{xx}^2\rangle_{k=0}}}
{(z^+_{opt}(k=0)-d_0)(z^-_{opt}(k=0)-d_0)}~,
\end{equation}
and $\sigma_{opt}(k)$ and $\omega_{opt}(k)$ in (\ref{zopt}) are the damping and frequency 
of the optic modes, which 
 have the following nonzero values in the long-wavelength limit:
\begin{equation}\label{z0opt}
\begin{array}{l}
\sigma_0=\frac{\langle\omega_{xx}^2\rangle_{k=0} D_{12}S_{xx}}{2x_1x_2k_BT}\\
\omega_0=\sqrt{\langle\omega_{xx}^2\rangle_{k=0}-\sigma_0^2}
\end{array}
\end{equation}
It is easily to check the sum rule, that the sum of all four eigenvalues (\ref{z12rel}) and 
(\ref{zopt}) equals the sum of the diagonal matrix elements of ${\bf T}^{(4T)}(k)$.
Having the solutions for optic branch (\ref{zopt}) we can exclude them from the characteristic 
polynomial of the eigenvalue problem for $(4\times 4)$ matrix ${\bf T}^{(4T)}(k)$ (\ref{Thyd})
and obtain an effective quadratic equation 
\begin{equation}\label{quadr}
z^2-(c_T^2/D_v-\Delta a^2 k^2)z-k^2(x_1x_2a^2k^2
-c_T^2\langle\omega_{xx}^2\rangle)/\langle\omega_{xx}^2\rangle=0~,
\end{equation}
which immediately results in two eigenmodes 
\begin{equation}\label{2modes}
z_{1,2}(k)=\delta(k)\pm \sqrt{\delta^2(k)+k^2(\frac{x_1x_2a^2k^2}
{\langle\omega_{xx}^2\rangle}-c^2_T)}
\end{equation}
where $\delta(k)=(\frac{c_T^2}{D_v}-\Delta a^2k^2)/2$. Here the term with 
$\Delta$ appears due to coupling of the non-hysrodynamic shear-stress 
relaxation with optic modes. It is straightforward to check 
that when $k\to 0$ one obtains exactly the two relaxing modes (\ref{z1}) and (\ref{z2}).
The shear waves emerge in the binary liquid when the expression under square 
root becomes negative and $z_{1,2}(k)$ in (\ref{2modes}) become a pair of complex-conjugated 
numbers.  One can see, that the second term under square root in (\ref{2modes}) is 
a difference of a positive function of $k^4$ and  $c^2_Tk^2$. The latter 
is the square of "bare" dispersion of non-damped transverse sound-like 
modes with the propagation speed $c_T$, while the former renormalizes the 
"bare" dispersion down preventing appearence of damped shear waves especially 
for large values of $t-x$ cross-correlation coupling $a$. The equation 
for the gap for shear waves in binary liquids follows from the expression
under square root in (\ref{2modes}) and reads:
\begin{equation}\label{kgap}
(\frac{x_1x_2a^2}{\langle\omega_{xx}^2\rangle}+\frac{\Delta^2 a^4}{4})k_s^4-
(c^2_T+d_0\Delta a^2)k_s^2+\frac{d_0}{4}=0~,
\end{equation}
which is a biquadratic equation for the propagation gap $k_s$. One can easily 
make sure that when the $t-x$ cross-correlations are absent, i.e. $a=0$, one 
obtains exactly the expression (\ref{k-gap}) for $k_s$.
Hence, the effect of mass-concentration fluctuations on the gap for 
shear waves is defined by the $t-x$ coupling constant $a$ and 
parameter $\Delta$ (\ref{Delta}) which accounts for the difference between 
the non-hydrodynamic eigenmodes (shear-stress relaxation and optic 
modes) in the long-wavelength limit.

\section{Conclusion}

We performed a study of transverse collective excitations in four binary Kob-Andersen 
liquids at the same density $n^*=1.2$ and temperature $T^*=2.0$ but with different mass 
ratio R. The spectra of transverse collective excitations in binary liquids contain 
two branches: high-frequency one of transverse optic modes and low-frequency one of 
shear waves. A striking feature was observed in the behavior of the low-frequency branch:
 the width of propagation gap of shear waves increased with the mass ratio R. For very 
large mass ratio R there is the tendency of the disappearence of the low-frequency 
branch, that is in agreement with previous simulation and GCM results \cite{Mry09}.

The standard expression for the gap width for shear waves, derived for pure liquids
is not able to explain the onset of non-hydrodynamic transverse excitations in binary 
liquids especially at large mass ratio of components. We solved analytically 
a 4-variable dynamic model with accounting for "t-x" cross-correlations between total mass 
and mass-concentration current correlations. We derived a new equation for the gap width
of shear waves in binary liquids, which reduces to the known gap width expression in 
case of absent "t-x" cross-correlations.

Our findings have another consequence. They show that the dispersion of 
collective excitations outside the hydrodynamic regime is strongly dependent on the 
mass ratio R, even if the static structure is identical. This raises a question about the 
validity of the "phonon theory of liquid thermodynamics" \cite{Bol12}, which explains the leading 
contribution to the heat capacity $C_v$ as coming from the most short-wavelength longitudinal
and transverse non-damped excitations. According to the "phonon theory of liquid thermodynamics"
in the case of absent shear waves the heat capacity $C_v$ should not be higher than 2$k_B$.
However, in our simulations all four binary systems with identical static structure 
resulted in the heat capacity $C_v=2.70\pm 0.02~k_B$ 
that makes evidence of the independence of the heat capacity on the width of propagation gap 
of shear waves. Our simulation study of pure and binary lqiuids on the "phonon theory of 
liquid thermodynamics" will be published elsewhere.

\section*{Acknowledgment}
This study was supported by the National Research Foundation of Ukraine under 
the grant No. 2020.02/0115.

\end{document}